\documentclass[final,3pt,scale=0.8,times,12pt]{elsarticle}
\usepackage{CJK} 
\biboptions{numbers,sort&compress}
\usepackage{graphicx,enumitem} 
\usepackage[margin=0.75in]{geometry}
\usepackage{amssymb}
\usepackage{amsmath}
\usepackage{subfigure}
\usepackage{epstopdf}
\usepackage{bibentry}
\usepackage{caption}
\usepackage{color}
\usepackage{mathrsfs}
\usepackage{multirow}
\usepackage{makecell}
\usepackage{hhline}
\usepackage{breqn}
\usepackage{lineno,hyperref}
\usepackage{fancyhdr}
\usepackage{geometry}
\DeclareMathOperator{\sech}{sech}

\newtheorem{rk}{Remark}
\journal{Journal of Mathematical Physics}
\allowdisplaybreaks[4]

\def\e{{\rm e}}

\def\e{{\rm e}}

\begin{document}
\begin{frontmatter}

\title{\textbf{Localized stem structures in quasi-resonant two-soliton solutions for the asymmetric Nizhnik-Novikov-Veselov system}}

\author[a]{Feng Yuan}
\author[b]{Jiguang Rao}
\author[c]{Jingsong He\corref{cor}}
\author[d]{Yi Cheng}
\address[a]{College of Science, Nanjing University of Posts and Telecommunications, Nanjing, 210023, P. R. China}
\address[b]{School of Mathematics and Statistics, Hubei University of Science and Technology, Xianning 437100, P. R. China}
\address[c]{Institute for Advanced Study, Shenzhen University, Shenzhen, 518060, P. R. China}
\address[d]{School of Mathematical Sciences, USTC, Hefei, Anhui 230026, P. R. China}

\cortext[cor]{Corresponding author. E-mails: hejingsong@szu.edu.cn}

\begin{abstract}
Elastic collisions of solitons generally have a finite phase shift. When the phase shift has a finitely large value, the two vertices of the (2+1)-dimensional 2-soliton are significantly separated due to the phase shift, accompanied by the formation of a local structure connecting the two V-shaped solitons. We define this local structure as the stem structure. This study systematically investigates the localized stem structures between two solitons in the (2+1)-dimensional asymmetric Nizhnik-Novikov-Veselov system. These stem structures, arising from quasi-resonant collisions between the solitons, exhibit distinct features of spatial locality and temporal invariance. We explore two scenarios: one characterized by weakly quasi-resonant collisions (i.e. $a_{12}\approx 0$), and the other by strongly quasi-resonant collisions (i.e. $a_{12}\approx +\infty$). Through mathematical analysis, we extract comprehensive insights into the trajectories, amplitudes, and velocities of the soliton arms. Furthermore, we discuss the characteristics of the stem structures, including their length and extreme points. Our findings shed new light on the interaction between solitons in the (2+1)-dimensional asymmetric Nizhnik-Novikov-Veselov system.
\end{abstract}


\end{frontmatter}

\section{Introduction}
Due to their extensive applications in physics and engineering, the theory and experimental researches of nonlinear waves are flourishing. The intricate evolution of nonlinear wave packets is described by numerous nonlinear partial differential equations (NPDEs). Over the past few decades, these equations have been well studied analytically and numerically. A plethora of methods emerged for analyzing and dissecting solutions to various NPDEs, including the Inverse scattering method \cite{prl1967,ip2020}, the Darboux transformation\cite{dt01,dt02,dt03,dt04}, and the Hirota bilinear method \cite{book01,nn2022,rao01,rao02}. These techniques empowered researchers to tackle NPDEs more effectively, boosting the discoveries of a large number of nonlinear wave solutions such as breather, lump, rogue wave, and their hybrid solutions.

Among these solutions, solitons are the first to be studied. In 1965, Zabusky and Kruskal published a groundbreaking paper introducing the concept of soliton numerically for the Korteweg-de Vries (KdV) equation \cite{prl1965}. Shortly after that, the $n$-soliton solutions of the KdV equation were successfully derived in 1974 analytically by using newly invented tool--inverse scattering method \cite{cpam1974}. After that, the soliton solutions have been constructed in
many integrable partial differential equations (see a early collection in reference \cite{book03}). The potential of the soliton solutions in explaining natural phenomena is further explored and has been extended into various fields \cite{book04}, such as plasma, nonlinear optics, Bose-Einstein condensation and many other fields.

The study of soliton interactions is a fundamental aspect of soliton theory. Solitons typically exhibit the following characteristics \cite{prl1965,cpam1974,book03,book04,interaction03}:
\begin{itemize}
\setlength{\itemsep}{-5pt}
  \item Solitons are spatially localized traveling wave solutions, maintaining a constant shape and velocity throughout their motion.
  \item If interactions between multiple solitons are elastic, after collision, the solitons regain their initial velocities with an extra phase shift. The phase of a soliton refers to the position of its wave crest, while phase shift denotes the variation in the soliton's phase during transmission. This phase shift is primarily determined by the interplay between nonlinear effects and dispersive effects, maintaining constancy in both the time and frequency domains.
  \item Elastic collisions of two crossed solitons result in an X-shaped configuration, leading to their classification as X-shape solitons.
\end{itemize}

The elastic collisions display discernible finite phase shifts. Moreover, when the phase shift in an elastic collision approaches infinity, albeit remaining finite, it is termed a quasi-resonant collision. In recent years, heightened scholarly attention has been directed towards the investigation of high-dimensional soliton equations, particularly those in the (2+1)-dimensional framework. Ordinarily, two-dimensional line solitons propagate infinitely across space; however, notably, these collisions can engender localized stem structures. The concept of a stem structure originally pertains to the intermediary wave linking the incident and reflected waves in Mach reflection (see \cite{stem01,stem02}). Under the quasi-resonant state, the vertices of X-shaped solitons become substantially spaced apart due to phase delay, forming two V-shaped regions interconnected by a novel isolated wave, also referred to as a stem (see Ref.\ \cite{stem03}). In this paper, we define the local structure that connects different soliton vertices, generated by the interaction between solitons, as the stem structure.

The stem structure of solitons has been discussed in previous studies, though the research is not extensive. For instance, Reference \cite{jpsj1980} examined the construction of quasi-resonant solitons within an extended Boussinesq-like equation. Reference \cite{jpsj1983-1} provided asymptotic forms of quasi-resonant two-solitons for the Kadomtsev-Petviashvili equation, wherein the stem structure is referred to as a virtual soliton. In 2012, Mark J. Ablowitz and Douglas E. Baldwin \cite{interaction09} reported quasi-resonant two-soliton water waves observed near low tide on two flat beaches located approximately 2000 km apart. However, only a few graphs (Fig. 3 and 4 in Ref. \cite{interaction09}) were presented, and the local characteristics of the intermediate wave (stem structure), such as its height, length and location, were not subjected to further analysis.

Considering the prevailing researches mainly focus on the solitons themselves and the scant detailed explorations on the attributes of the stem structure beyond preliminary analysis and intuitive graphical representation, we aims to investigate the localized stem structures within the (2+1)-dimensional asymmetric Nizhnik-Novikov-Veselov (ANNV) system through analytical methods and delve into their local properties.

The (2+1)-dimensional asymmetric Nizhnik-Novikov-Veselov (ANNV) system, first introduced by Boiti et al., is represented by the following form \cite{annv01}:
\begin{equation}\label{annveq}
	\left\{
	\begin{aligned}
		&u_t+v_{xxx}=3(uv)_x,\\
		&u_x=v_y.\\
	\end{aligned}
	\right.
\end{equation}
Here, $u$ and $v$ represent the components of dimensionless velocity. The ANNV system \eqref{annveq} can explain various important physical phenomena such as the shallow waves driven by weakly nonlinear restoring forces in incompressible fluids, the long internal waves within a density-stratified ocean, and the acoustic waves on a crystal lattice. The ANNV system extends the scope of the Hirota and Satsuma equation \cite{annv03}, making it a versatile framework for understanding diverse physical systems. Reference \cite{annv01} proved that Eq.\ \eqref{annveq} can be simplified as the KdV equation when $x=y$ \cite{annv01}. It can be also derived through the application of the inner parameter-dependent symmetry constraint of the KP equation \cite{annv04}. Clarkson and Mansfield have explored several key aspects of the ANNV system, including the examination of the Painlev\'{e} property and the investigation of similarity solutions \cite{annv05}. Additionally, various types of solutions have been reported, including dromion and kink solutions \cite{annv02,annv16}, variable separation solutions \cite{annv06}, quasi-periodic solutions \cite{annv07}, soliton solutions \cite{annv10}, lump-type solutions \cite{annv08,annv11,annv12} and rational and semi-rational solutions \cite{annv09,annv15}.

Recently, resonance Y-shape soliton solutions of Eq.\ \ref{annveq} have been formulated in Ref. \cite{annv13,annv14}. However, the X-shape solitons with constant length stems have not been constructed. The essential difficult problem in this study is to determine explicit  expressions of the stem wave and its two ends. This leaves considerable room for understanding soliton interactions, which forms the primary focus of this paper. The basic purpose of us is to overcome unravelled problem in references \cite{interaction09,annv13,annv14} on the stem structure and then establish following objectives: (1) We construct constant length stem structure in two soliton solutions generated by quasi-resonant collisions. (2) We further propose two distinct ways of soliton construction, namely, via the weakly quasi-resonant collision when $a_{12}\approx 0$ and the strongly quasi-resonant collision when $a_{12}\approx +\infty$. (3) Meanwhile, we, for the first time, provide systematical exploration on the localization and dynamical characteristics of the stem structures.

The structure of the paper is organized as follows: In Sec. \ref{sec2}, we introduce the expressions of the Hirota Bilinear form and solutions of the ANNV system \eqref{annveq}. In Sec. \ref{sec3} and \ref{sec4}, we investigate the constant length stem structures in two soliton solutions generated by the weakly and strongly quasi-resonant collision, respectively. Their localization and dynamical properties are also provided. Finally, in Sec. \ref{summary}, we summarize and discuss our results.
\section{Basic formulas of Hirota Bilinear form and soliton solutions}\label{sec2}
In this section, we recall the Hirota Bilinear form and $n$-soliton solutions of the ANNV system \eqref{annveq}. By using the transformation
\begin{flalign}\label{uv}
	u=-2(\ln f)_{xx},\,v=-2(\ln f)_{xy},
\end{flalign}
it has the bilinear form\cite{jmp1994}
\begin{equation}\label{bilinear}
D_y(D_t+D_x^3)f\cdot f=0,
\end{equation}
where $D$ is defined by \cite{book01,book02}
\begin{flalign*}
	D_t^mD_x^nD_y^r f(x,y,t)\cdot g(x,y,t)=\Big(\frac{\partial}{\partial t}-\frac{\partial}{\partial t'}\Big)^m\Big(\frac{\partial}{\partial x}-\frac{\partial}{\partial x'}\Big)^n\Big(\frac{\partial}{\partial y}-\frac{\partial}{\partial y'}\Big)^rf(x,y,t)g(x',y',t')\Big|_{x'=x,y'=y,t'=t},
\end{flalign*}
and $m,\,n,\,r$ are nonnegative integers. The N-soliton solutions of \eqref{annveq} have been given by the following form:
\begin{equation}\label{nf}
	f^{[N]}=\sum_{\mu =0, 1} \exp \left(\sum_{i<j}^{N} \mu_{i}\mu_{j}A_{ij}+ \sum_{i=1}^{N} \mu_{i}\xi_{i}\right),
\end{equation}
where,
\begin{flalign}\label{etakpw}
	\begin{split}
		\xi_j=k_jx+p_jy-k_j^3t+\xi_j^0,\,\exp(A_{ij})=\frac{(k_i-k_j)(p_i-p_j)}{(k_i+k_j)(p_i+p_j)}\triangleq a_{ij}\geqslant 0.
	\end{split}
\end{flalign}

The 1-soliton solution of Eq. \eqref{annveq} is $u^{[1]}=-\frac{k_1p_1}{2}\sech^2(\frac{\xi_1}{2}),\,v^{[1]}=-\frac{k_1^2}{2}\text{sech}^2(\frac{\xi_1}{2})$. It's readily to know that both $u^{[1]}$ and $v^{[1]}$ have trajectories of $\xi_1=0$, while their amplitudes are $u_{amp}=-\frac{k_1p_1}{2}$ and $v_{amp}=-\frac{k_1^2}{2}$, respectively. Then it can be seen that $u^{[1]}$ is a dark soliton when $k_1p_1>0$ and it is a bright soliton when $k_1p_1<0$, while $v^{[1]}$ is always a dark soliton.

In this paper, we focus on the 2-soliton solution of the ANNV system. Substituting $N=2$ into Eq. \eqref{nf}, the fundamental 2-soliton solution can be obtained by \eqref{uv} and the following equation,
\begin{flalign}\label{2soliton}
	\begin{split}
		f^{[2]}=1+e^{\xi_1}+e^{\xi_2}+a_{12}e^{\xi_1+\xi_2}.
	\end{split}
\end{flalign}

The phase shift of the 2-soliton solution is denoted as $\Delta_{12}=\ln a_{12}$. Different conditions on the phase shift give rise to distinct types of collisions between the two solitons, such as elastic collisions and resonance collisions. In the context of elastic collisions, a 2-soliton is X-shaped. When the phase shift is sufficiently large but finite, the two vertices of the X-shaped soliton are significantly separated due to this phase shift, forming a local structure that connects the two V-shaped solitons. Since 2-solitons are travelling waves, the stem structure also maintains its shape during propagation, resulting in a constant-length stem structure. The constant-length stem structure in quasi-resonant 2-solitons is the primary focus of our investigation. Subsequently, we will delve into the analysis of the localized stem structure under the condition $\Delta_{12}\approx\infty$, corresponding to either $a_{12}\approx 0$ or $a_{12}\approx +\infty$. We will explore these two cases in detail.

\begin{rk}
The distinction between strongly and weakly resonances lies in their outcomes: a strongly resonance (\(a_{ij}=+\infty\)) between \(S_i\) (\(f_i=1+\exp(\xi_i)\)) and \(S_j\) (\(f_j=1+\exp(\xi_j)\)) produces a soliton \(S_{i+j}\) (\(f_{i+j}=1+\exp(\xi_i+\xi_j)\)), whereas a weakly resonance (\(a_{ij}=0\)) yields \(S_{i-j}\) (\(f_{i+j}=1+\exp(\xi_i-\xi_j)\)). Similarly, we refer to strongly quasi-resonance when \(a_{ij}\approx +\infty\) and weakly quasi-resonance when \(a_{ij}\approx 0\).
\end{rk}
\section{Constant length stem structure generated by weakly quasi-resonant collision}\label{sec3}
In the scenario where $a_{12}\approx 0$ ($\Delta_{12}\approx -\infty$), we initially examine the asymptotic properties of the 2-soliton by the asymptotic analysis method given in Ref.\ \cite{jpsj1980,jpsj1983-1,guo2023}. To analyze the asymptotic behavior of the soliton $S_1$, we express the tau function \eqref{2soliton} as:
\begin{flalign}\label{f01}
\begin{split}
f=1+e^{\xi_1}+e^{\xi_2}(1+e^{\xi_1+\ln a_{12}}).
\end{split}
\end{flalign}
\begin{table}
  \centering
  \caption{Physical quantities of the soliton arms}
  \label{tab:t1}
  \begin{tabular}{cccccc}
    \Xhline{1pt}
     Soliton & Trajectory & Velocity & Amplitude & Components \\
    \hline
    \multirow{2}{*}{$S_{j}$} & \multirow{2}{*}{$l_j$} & \multirow{2}{*}{$(k_j^2, -\frac{k_j}{p_j})$} & $-\frac{k_jp_j}{2}$ & $u_{j}$ \\
     & & &$-\frac{k_j^2}{2}$ & $v_{j}$ \\
    \hline
    \multirow{2}{*}{$S_{j}$} & \multirow{2}{*}{$\widetilde{l_j}$} & \multirow{2}{*}{$(k_j^2, -\frac{k_j}{p_j})$} & $-\frac{k_jp_j}{2}$ & $\widetilde{u_{j}}$ \\
     & & &$-\frac{k_j^2}{2}$ & $\widetilde{v_{j}}$ \\
     \hline
    \multirow{2}{*}{$S_{1-2}$} & \multirow{2}{*}{$l_{1-2}$} & \multirow{2}{*}{$(k_1^2+k_1k_2+k_2^2,\,-\frac{k_1^3-k_2^3}{p_1-p_2})$} & $-\frac{(k_1-k_2)(p_1-p_2)}{2}$ & $u_{1-2}$ \\
     & & &$-\frac{(k_1-k_2)^2}{2}$ & $v_{1-2}$ \\
     \Xhline{1pt}
  \end{tabular}
  \caption*{\captionsetup{justification=raggedright,singlelinecheck=false,format=hang} \quad The $j$-th soliton $S_j$ $(j=1,2,1-2)$ is composed by two components, $u_j$ and $v_j$, and their key properties are summarized by Table \ref{tab:t1}. The relevant formulas are listed by \eqref{l12}.}
\end{table}

For $\xi_1\approx 0,\,\xi_2\to -\infty$, Eq. \eqref{f01} can be approximated by
\begin{flalign*}
\begin{split}
f\approx f_1=1+e^{\xi_1}.
\end{split}
\end{flalign*}
Similarly, for $\xi_1+\ln a_{12}\approx 0,\,\xi_2\to +\infty$, Eq. \eqref{f01} can be approximated by
\begin{flalign*}
\begin{split}
f\approx \widetilde{f_1}=1+a_{12}e^{\xi_1}.
\end{split}
\end{flalign*}

The asymptotic properties of soliton $S_2$ are studied in a similar manner, with the tau function \eqref{2soliton} expressed by
\begin{flalign}\label{f02}
\begin{split}
f=1+e^{\xi_2}+e^{\xi_1}(1+e^{\xi_2+\ln a_{12}}).
\end{split}
\end{flalign}

For $\xi_2\approx 0,\,\xi_1\to -\infty$, Eq. \eqref{f02} can be approximated by
\begin{flalign*}
\begin{split}
f\approx f_2=1+e^{\xi_2}.
\end{split}
\end{flalign*}

For $\xi_2+\ln a_{12}\approx 0,\,\xi_1\to +\infty$, Eq. \eqref{f01} can be approximated by
\begin{flalign*}
\begin{split}
f\approx \widetilde{f_2}=1+a_{12}e^{\xi_2}.
\end{split}
\end{flalign*}

Finally, we examine the asymptotic properties of the constant length stem. In the case of $\xi_1\approx \xi_2,\,\xi_{1,2}\to +\infty$, and $a_{12}\approx 0$, we observe that $e^{\xi_1}+e^{\xi_2}$ is much larger than $1+a_{12}e^{\xi_1+\xi_2}$. Consequently, Eq. \eqref{f02} can be approximated as:
\begin{flalign*}
\begin{split}
f\approx f_{1-2}=e^{\xi_1}+e^{\xi_2}=e^{\xi_2}(1+e^{\xi_1-\xi_2}).
\end{split}
\end{flalign*}

Based on the above asymptotic analysis, the 2-soliton undergoes weakly quasi-resonant collisions, as manifested in the following asymptotic forms:\\
Before collision:
\begin{flalign}
\begin{split}
\text {The soliton } &S_1\,(\xi_1\approx 0,\,\xi_2\to -\infty):\,u\approx u_{1},\,v\approx v_{1},\\
\text {The soliton } &S_2\,(\xi_2+\ln a_{12}\approx 0,\,\xi_1\to +\infty):\,u\approx \widetilde{u_{2}},\,v\approx \widetilde{v_{2}},\\
\end{split}\label{xasy01}
\end{flalign}
After collision:
\begin{flalign}
\begin{split}
\text {The soliton } &S_1\, (\xi_1+\ln a_{12}\approx 0,\,\xi_2\to +\infty):\,u\approx \widetilde{u_{1}},\,v\approx \widetilde{v_{1}},\\
\text {The soliton } &S_2\,(\xi_2\approx 0,\,\xi_1\to -\infty):\,u\approx u_{2},\,v\approx v_{2};
\end{split}\label{xasy02}
\end{flalign}
The constant length stem:
\begin{flalign}
	\begin{split}
\text {The soliton } &S_{1-2}\, (\xi_1\approx \xi_2,\,\xi_{1,2}\to +\infty):\,u\approx u_{1-2},\,v\approx v_{1-2}.
	\end{split}\label{xasy03}
\end{flalign}

Here, the formulas of the soliton arms in \eqref{xasy01}--\eqref{xasy03} are given as following,
\begin{flalign}\label{uv01}
	\begin{split}
&u_{j}=-\frac{k_jp_j}{2}\sech^2(\frac{\xi_j}{2}),\,v_{j}=-\frac{k_j^2}{2}\sech^2(\frac{\xi_j}{2}),\,j=1,\,2,\\
&\widetilde{u_{j}}=-\frac{k_jp_j}{2}\sech^2(\frac{\xi_j+\ln a_{12}}{2}),\,\widetilde{v_{j}}=-\frac{k_j^2}{2}\sech^2(\frac{\xi_j+\ln a_{12}}{2}),\\
&u_{1- 2}=-\frac{(k_1- k_2)(p_1- p_2)}{2}\sech^2(\frac{\xi_1- \xi_2}{2}),\,v_{1- 2}=-\frac{(k_1- k_2)^2}{2}\sech^2(\frac{\xi_1- \xi_2}{2}).
	\end{split}
\end{flalign}

The stem $S_{1-2}$ is also denoted as a virtual soliton and was initially introduced in reference \cite{jpsj1980} for the extended Boussinesq-like equation, and subsequently in reference \cite{jpsj1983-1} for the Kadomtsev-Petviashvili equation. These structures have been depicted graphically. It is noteworthy that the localized characteristics of the stem structures, including trajectory and endpoint coordinates, have not been rigorously analyzed. Consequently, our attention now shifts towards the analytical study concerning the stem structure.

Undoubtedly, this soliton exhibits five arms. Table \ref{tab:t1} provides the formulas, trajectories, amplitudes, and velocities for each arm, while the pertinent formulae are given by \eqref{uv01} and
\begin{flalign}\label{l12}
\begin{split}
&\boldsymbol{l_1:}\,\xi_1=0,\,\quad\,\boldsymbol{l_2:}\, \xi_2=0,\,\quad \boldsymbol{\widetilde{l_1}:}\,\xi_1+\ln a_{12}=0,\,\quad \boldsymbol{\widetilde{l_2}:}\, \xi_2+\ln a_{12}=0,\,\quad \boldsymbol{l_{1-2}:}\, \xi_1-\xi_2=0.
\end{split}
\end{flalign}

\begin{figure}[htbp]
	\centering
    \subfigure[$u$]{\includegraphics[height=3.7cm,width=3.7cm]{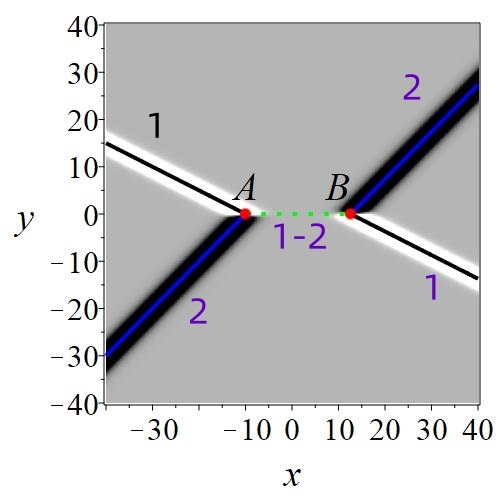}}
    \subfigure[$v$]{\includegraphics[height=3.7cm,width=3.7cm]{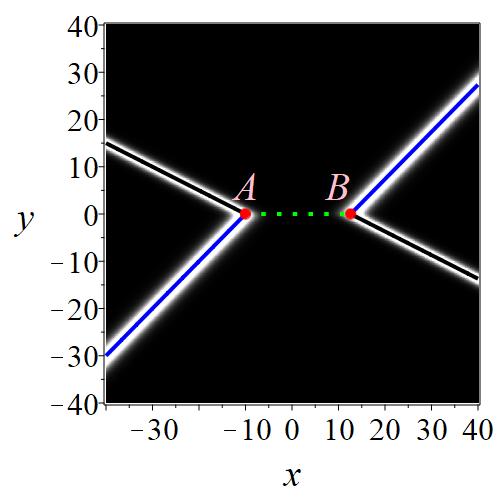}}
	\subfigure[$u$]{\includegraphics[height=3.7cm,width=3.7cm]{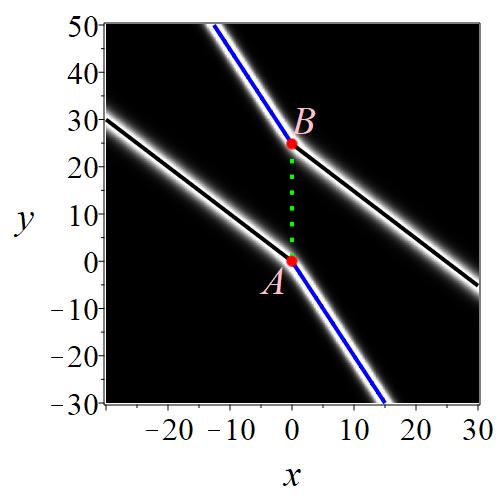}}
	\subfigure[$v$]{\includegraphics[height=3.7cm,width=3.7cm]{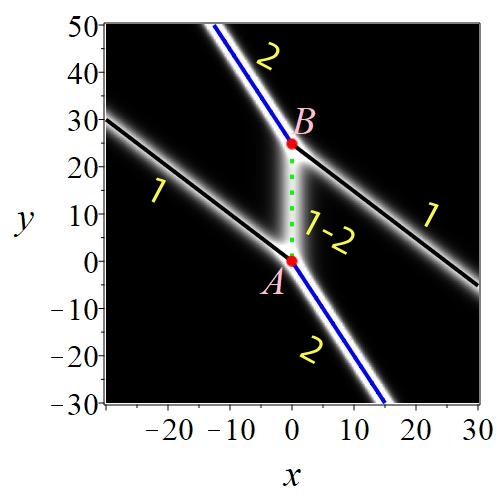}}
	\caption{The density plots of the quasi-resonant two soliton with constant length stem structure. Parameters: (a) (b) $k_1=1,\,k_2=1-10^{-10},\,p_1=2,\,p_2=-1,\,\xi_1^0=10,\,\xi_2^0=10,\,t=0$. (c) (d) $k_1=1,\,k_2=2,\,p_1=1,\,p_2=1+10^{-10},\,\xi_1^0=0,\,\xi_2^0=0,\,t=0$.  The lines are the trajectories of $S_j$, and the red points are the endpoints of the stem which are expressed as Eqs.\ \eqref{eqa} and \eqref{eqb}.}\label{fig1-1}
\end{figure}

Solving a group of equations $\xi_1=0$ and $\xi_2=0$ implies an intersection point $A$ on ($x,y$)-plane of $l_1$ and $l_2$ as:
{\small\begin{flalign}
\begin{split}\label{eqa}
A\, \left(\frac{(k_1^3p_2-k_2^3p_1)t+p_1\xi_2^0-p_2\xi_1^0}{k_1p_2-k_2p_1},\,-\frac{(k_1^3k_2-k_1k_2^3)t+k_1\xi_2^0-k_2\xi_1^0}{k_1p_2-k_2p_1} \right).
\end{split}
\end{flalign}}
Similarly, the intersection point $B$ on ($x,y$)-plan of $\widetilde{l_1}$ and $\widetilde{l_2}$ can be generated as
{\small\begin{flalign}
\begin{split}\label{eqb}
B\, \left(\frac{(p_1-p_2)\ln a_{12}+(k_1^3p_2-k_2^3p_1)t+p_1\xi_2^0-p_2\xi_1^{0}}{k_1p_2-k_2p_1},\,-\frac{(k_1-k_2)\ln a_{12}+(k_1^3k_2-k_1k_2^3)t+k_1\xi_2^0-k_2\xi_1^0}{k_1p_2-k_2p_1} \right),
\end{split}
\end{flalign}}
by solving a group of $\xi_1+\ln a_{12}=0$ and $\xi_2+\ln a_{12}=0$. It is noteworthy that points $A$ and $B$ also serve as the endpoints of $l_{1-2}$ , which can be seen in Fig. \ref{fig1-1}. Consequently, the length of the stem, denoted as $|AB|$, is defined as:
\begin{equation}\label{eqab}
|AB| = \left| \frac{\ln a_{12}}{k_1p_2-k_2p_1} \right| \sqrt{(k_1-k_2)^2+(p_1-p_2)^2}.
\end{equation}

This formula show that the length of the stem is constant. To ensure $a_{12}\approx 0$, it is imperative to set $k_1\approx k_2$ or $p_1\approx p_2$. Specifically, if $k_1\approx k_2$, the stem's amplitude is almost zero. Consequently, the segment between $A$ and $B$ in Fig.\ \ref{fig1-1} merges with the background plane. However, it is crucial to recognize that this segment comprises solitons of exceptionally small amplitude, as depicted in Fig.\ \ref{fig1-1} (a) and (b). Notably, when $p_1p_2<0$, $u$ manifests as a dark-bright soliton, while $p_1>0$ and $p_2>0$ make $u$ a 2-bright soliton. Conversely, if $p_1\approx p_2$, the amplitude of the stem $u_{1-2}$ nearly vanishes, while $v_{1-2}$ remains non-zero, as illustrated in Fig.\ \ref{fig1-1} (c) and (d). In this scenario, $k_1>0$ and $k_2>0$ make $u$ a 2-dark soliton, whereas $k_1k_2<0$ results in $u$ being a dark-bright soliton.

Next, we explore the cross-sectional curves of the 2-soliton \eqref{uv} on the planes $\xi_1-\xi_2=0$:
\begin{flalign}\label{cross21}
\begin{split}
\text{When $k_1\approx k_2$,}&\\
u|_{l_{1-2}}^{(1)}=&-\frac{2(k_1p_1+k_2p_2)g_1g_2\e^{3\theta_1}+4g_1^2g_2\e^{2\theta_1}+2(k_1p_1+k_2p_2)g_1^2\e^{\theta_1}}{(g_2\e^{2\theta_1}+2g_1\e^{\theta_1}+g_1)^2};\\
v|_{l_{1-2}}^{(1)}=&-\frac{2(k_1^2+k_2^2)g_1g_2\e^{3\theta_1}+4(k_1p_1-k_2p_2)(k_1^2-k_2^2)g_1e^{2\theta_1}+2g_1^2(k_1^2+k_2^2)\e^{\theta_1}}
{(g_2\e^{2\theta_1}+2g_1\e^{\theta_1}+g_1)^2};
\end{split}
\end{flalign}
\begin{flalign}\label{cross22}
\begin{split}
\text{When $p_1\approx p_2$,}&\\
u|_{l_{1-2}}^{(2)}=&-\frac{2(k_1p_1+k_2p_2)g_1g_2\e^{3\theta_2}+4g_1^2g_2\e^{2\theta_2}+2(k_1p_1+k_2p_2)g_1^2\e^{\theta_2}}{(g_2\e^{2\theta_2}+2g_1\e^{\theta_2}+g_1)^2};\\
v|_{l_{1-2}}^{(2)}=&-\frac{2(k_1^2+k_2^2)g_1g_2\e^{3\theta_2}+4(k_1p_1-k_2p_2)(k_1^2-k_2^2)g_1e^{2\theta_2}+2g_1^2(k_1^2+k_2^2)\e^{\theta_2}}
{(g_2\e^{2\theta_2}+2g_1\e^{\theta_2}+g_1)^2};
\end{split}
\end{flalign}

Here,
\begin{flalign}\label{g12}
\begin{split}
&\theta_1=\frac{p_1(k_2x-k_2^3t+\xi_2^0)-p_2(k_1x-k_1^3t+\xi_1^0)}{p_1-p_2},\,\theta_2=\frac{k_1(p_2y-k_2^3t+\xi_2^0)-k_2(p_1y-k_1^3t+\xi_1^0)}{k_1-k_2},\,\\
&g_1=(p_1+p_2)(k_1+k_2),\quad g_2=(p_1-p_2)(k_1-k_2).
\end{split}
\end{flalign}

\begin{figure}[h!tb]
	\centering
	\raisebox{18ex}{(a)}\includegraphics[height=4cm,width=3.8cm]{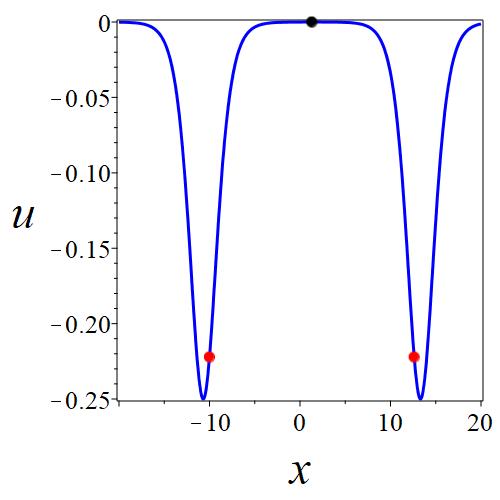}
	\raisebox{18ex}{(b)}\includegraphics[height=4cm,width=3.8cm]{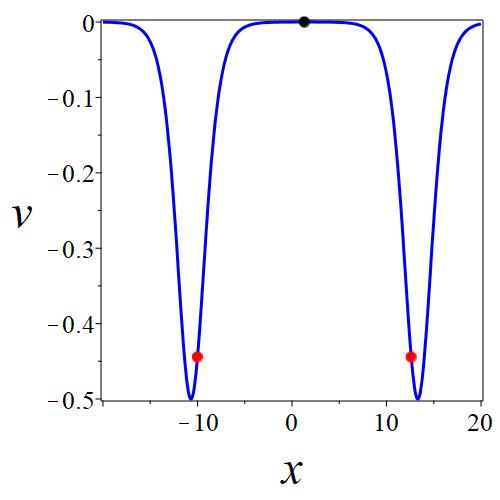}
    \raisebox{18ex}{(c)}\includegraphics[height=4cm,width=3.8cm]{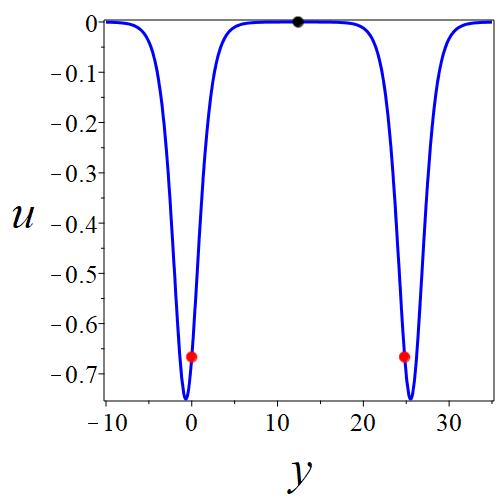}
	\raisebox{18ex}{(d)}\includegraphics[height=4cm,width=3.8cm]{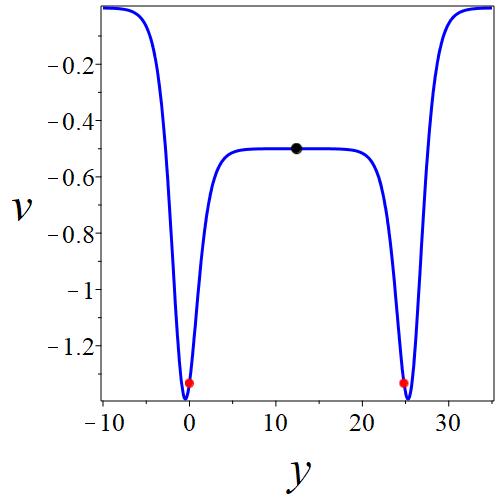}
	\caption{(a) (b) The cross-sectional curves \eqref{cross21} with parameters: $k_1=1,\,k_2=1-10^{-10},\,p_1=2,\,p_2=-1$; The black points are corresponding to $P_1$, and the red points are corresponding to $A$ and $B$. (c) (d) The cross-sectional curves \eqref{cross22} with parameters: $k_1=1,\,k_2=2,\,p_1=1,\,p_2=1+10^{-10}$.
	}\label{fig1-2}
\end{figure}
\begin{figure}[h!tb]
	\centering
	\raisebox{18ex}{(a)}\includegraphics[height=4cm,width=3.8cm]{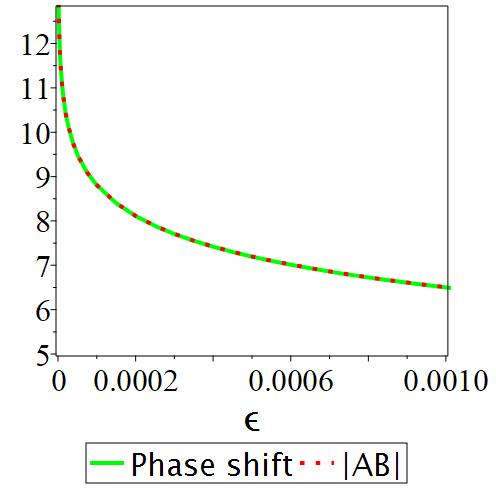}
	\raisebox{18ex}{(b)}\includegraphics[height=4cm,width=3.8cm]{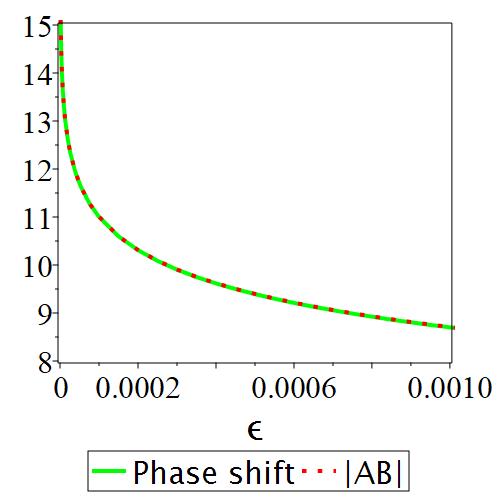}
	\caption{Parameters: (a) $k_1=1,\,k_2=1-\epsilon,\,p_1=2,\,p_2=-1$; (b) $k_1=1,\,k_2=2,\,p_1=1,\,p_2=1+\epsilon$.
	}\label{fig1-3}
\end{figure}

In general, the stem has a flat top on (or lower than) $(x,\,y)$-plane with two deep downward peaks, as verified in Fig.\ \ref{fig1-2} illustrating the cross-sectional curves corresponding to Fig.\ \ref{fig1-1}. Deriving the extreme values by taking the derivative of Eq.\ \eqref{cross21}, we observe that instead of a line soliton having an extreme value line, the constant length stem $S_{1-2}$ possesses only one extreme point between $A$ and $B$. The extreme points of $u|_{l_{1-2}}^{(1)}$ and $v|_{l_{1-2}}^{(1)}$ share the same coordinates on the $(x, y)$-direction, as follows:

{\small
\begin{flalign}\label{extrem1}
\begin{split}
P_1:\quad\left(\frac{(p_1-p_2)\ln a_{12}-2p_1(k_2^3t-\xi_2^0)+2p_2(k_1^3t-\xi_1^0)}{2(k_1p_2-k_2p_1)},\,-\frac{(k_1-k_2)\ln a_{12}-2k_1(k_2^3t-\xi_2^0)+2k_2(k_1^3t-\xi_1^0)}{2(k_1p_2-k_2p_1)}\right)
\end{split}
\end{flalign}}

\begin{rk}
Since 2-soliton propagates within the $(x,\,y)$-plane, the point \(P_1\) (the midpoint of the stem structure) also traverses the $(x,\,y)$-plane. Consequently, the coordinates of \(P_1\) are functions of time \(t\). This behavior similarly applies to the point \(P_2\) (as described in Eq.\ \eqref{extrem2} in Section \ref{sec4}).
\end{rk}

It is noteworthy that $P_1$ precisely corresponds to the midpoint of $AB$, represented by black dots in Fig.\ \ref{fig1-2}. Substituting \eqref{extrem1} into \eqref{uv} and \eqref{2soliton}, we obtain the extreme values of $S_{1-2}$ as $u(P_1)=-\frac{(k_1-k_2)(p_1-p_2)}{2\sqrt{a_{12}}}\approx -\frac{(k_1-k_2)(p_1-p_2)}{2}$ and $v(P_1)=-\frac{(k_1-k_2)^2}{2}\cdot\frac{(k_1+k_2)^2\sqrt{a_{12}}+(k_1-k_2)^2}{(k_1-k_2)^2(\sqrt{a_{12}}+1)}\approx -\frac{(k_1-k_2)^2}{2}$. These approximations further validate the accuracy of the asymptotic forms \eqref{xasy03}. We can see  from Fig. \ref{fig1-2} that the two ends of the stem are not the maximum or minimum points of $u$ or $v$.

Setting $|k_1-k_2|=\epsilon\approx0$ or $|p_1-p_2|=\epsilon\approx0$ with $|k_j|,|p_j|\gg\epsilon$, we find $|AB|\approx |\Delta_{12}|$. Figure \ref{fig1-3} depict how ``$|AB|$" and ``$|\Delta_{12}|$" change concerning ``$\epsilon$". It is evident that as $\epsilon\approx0$, the ``$|AB|$" curve closely aligns with the phase curve, and a smaller $\epsilon$ corresponds to a longer stem.
\section{Constant length stem structure generated by strongly quasi-resonant collision}\label{sec4}
In situations where $a_{12}\approx \infty,(\Delta_{12}\approx+\infty)$, the 2-soliton undergoes strongly quasi-resonant collisions. The asymptotical properties of the soliton $S_1$ and $S_2$ are the same as the case $a_{12}\approx 0$, but the asymptotical properties of the stem are different. In the case of $\xi_1\approx -\xi_2,\,\xi_1\to +\infty,\,\xi_2\to -\infty$, and $a_{12}\approx +\infty$, we observe that $1+a_{12}e^{\xi_1+\xi_2}$ is much larger than $e^{\xi_1}+e^{\xi_2}$. Consequently, Eq. \eqref{f02} can be approximated as:
\begin{flalign*}
\begin{split}
f\approx \widetilde{f_{1+2}}=1+a_{12}e^{\xi_1+\xi_2}.
\end{split}
\end{flalign*}

Based on the asymptotic analysis of $S_1,\,S_2$, and $S_{1+2}$, the 2-soliton experiences strongly elastic quasi-resonant manifests
in the following asymptotic forms:\\
Before collision:
\begin{flalign}
	\begin{split}
\text {The soliton } &S_1\,(\xi_1\approx 0,\,\xi_2\to -\infty):\,u\approx u_{1},\,v\approx v_{1},\\
\text {The soliton } &S_2\,(\xi_2+\ln a_{12}\approx 0,\,\xi_1\to +\infty):\,u\approx \widetilde{u_{2}},\,v\approx \widetilde{v_{2}};
\end{split}\label{xasy04}
\end{flalign}
After collision:
\begin{flalign}
	\begin{split}
\text {The soliton } &S_1\, (\xi_1+\ln a_{12}\approx 0,\,\xi_2\to +\infty):\,u\approx \widetilde{u_{1}},\,v\approx \widetilde{v_{1}},\\
\text {The soliton } &S_2\,(\xi_2\approx 0,\,\xi_1\to -\infty):\,u\approx u_{2},\,v\approx v_{2};
\end{split}\label{xasy05}
\end{flalign}
The constant length stem:
\begin{flalign}
	\begin{split}
\text {The soliton } S_{1+2}\, (\xi_1\approx -\xi_2,\,\xi_1\to +\infty,\,\xi_2\to -\infty):\,u\approx \widetilde{u_{1+2}},\, v\approx \widetilde{v_{1+2}}.
	\end{split}\label{xasy06}
\end{flalign}
The relevant expressions are provided in Eq.\ \eqref{uv01} and
 \begin{flalign}\label{uv02}
	\begin{split}
&\widetilde{u_{1+ 2}}=-\frac{(k_1+ k_2)(p_1+ p_2)}{2}\sech^2(\frac{\xi_1+ \xi_2+\ln a_{12}}{2}),\,\widetilde{v_{1+ 2}}=-\frac{(k_1+ k_2)^2}{2}\sech^2(\frac{\xi_1+ \xi_2+\ln a_{12}}{2}).
	\end{split}
\end{flalign}

This soliton also has five arms. The formulas, trajectories, amplitudes, velocities of these five arms before and after collision are provided in tables \ref{tab:t1} and \ref{tab:t2}, while the pertinent formulae are given by \eqref{l12} and following formula,
\begin{equation}\label{l3}
\boldsymbol{\widetilde{l_{1+2}}:}\, \xi_1+\xi_2+\ln a_{12}=0.
\end{equation}

Solving the system of equations $\xi_1=0$ and $\xi_2+\ln a_{12}=0$ leads to an intersection point $C$ on the ($x,\,y$)-plane of $l_1$ and $\widetilde{l_2}$:
{\small\begin{flalign}
\begin{split}\label{eqc}
C\, \left(\frac{p_1\ln a_{12}+(k_1^3p_2-k_2^3p_1)t+p_1\xi_2^0-p_2\xi_1^0}{k_1p_2-k_2p_1},\,-\frac{k_1\ln a_{12}+(k_1^3k_2-k_1k_2^3)t+k_1\xi_2^0-k_2\xi_1^0}{k_1p_2-k_2p_1}\right).
\end{split}
\end{flalign}}
Similarly, solving the system $\xi_1+\ln a_{12}=0$ and $\xi_2=0$ yields an intersection point $D$ on the ($x,y$)-plane of $l_2$ and $\widetilde{l_1}$:
{\small\begin{flalign}
\begin{split}\label{eqd}
D\, \left(\frac{-p_2\ln a_{12}+(k_1^3p_2-k_2^3p_1)t+p_1\xi_2^0-p_2\xi_1^0}{k_1p_2-k_2p_1},\,-\frac{-k_2\ln a_{12}+(k_1^3k_2-k_1k_2^3)t+k_1\xi_2^0-k_2\xi_1^0}{k_1p_2-k_2p_1}\right).
\end{split}
\end{flalign}}

\begin{table}
  \centering
  \caption{Physical quantities of the stem structures}
  \label{tab:t2}
  \begin{tabular}{cccccc}
    \Xhline{1pt}
    Stem & Trajectory & Velocity & Amplitude & Components \\
    \hline
    \multirow{2}{*}{$S_{1+2}$}  & \multirow{2}{*}{$\widetilde{l_{1+2}}$} & \multirow{2}{*}{$(k_1^2-k_1k_2+k_2^2,\,-\frac{k_1^3+k_2^3}{p_1+p_2})$} & $-\frac{(k_1+k_2)(p_1+p_2)}{2}$ & $\widetilde{u_{1+2}}$\\
     & & &$-\frac{(k_1+k_2)^2}{2}$ & $\widetilde{v_{1+2}}$ \\
    \Xhline{1pt}
  \end{tabular}
  \caption*{\captionsetup{justification=raggedright,singlelinecheck=false,format=hang} \quad The solitons $S_{1+2}$ is composed by two components $u_{1\pm 2}$ and $v_{1\pm 2}$, and their trajectories are listed by \eqref{l3}.}
\end{table}

Then we can also obtain the length of the constant length stem as
\begin{equation}\label{eqcd}
|CD|=\left| \frac{\ln a_{12}}{k_1p_2-k_2p_1} \right|\sqrt{(k_1+k_2)^2+(p_1+p_2)^2}.
\end{equation}

\begin{figure}[h!tb]
	\centering
    \subfigure[$u$]{\includegraphics[height=3.7cm,width=3.7cm]{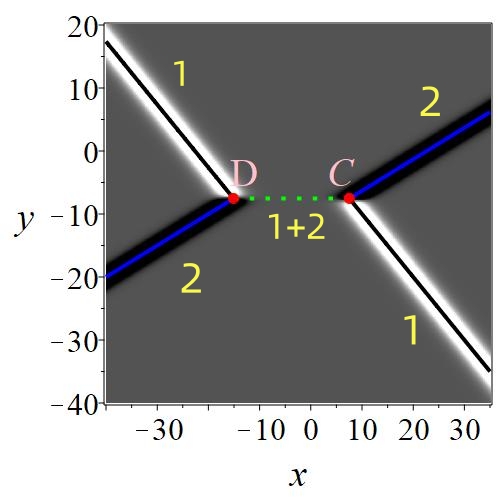}}
    \subfigure[$v$]{\includegraphics[height=3.7cm,width=3.7cm]{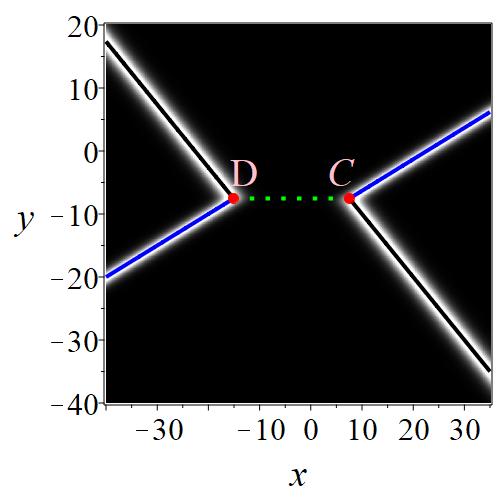}}
	\subfigure[$u$]{\includegraphics[height=3.7cm,width=3.7cm]{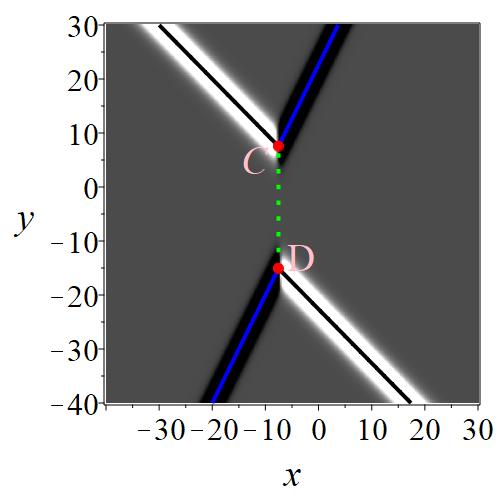}}
	\subfigure[$v$]{\includegraphics[height=3.7cm,width=3.7cm]{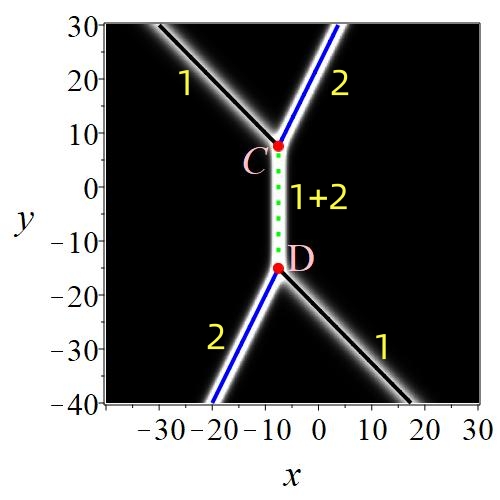}}
	\caption{The density plots of the quasi-resonant two soliton with constant length stem structure. Parameters: (a) (b) $k_1=1,\,k_2=-1-10^{-10},\,p_1=1,\,p_2=2,\,\xi_1^0=0,\,\xi_2^0=0,\,t=0$. (c) (d) $k_1=1,\,k_2=2,\,p_1=1,\,p_2=-1-10^{-10},\,\xi_1^0=0,\,\xi_2^0=0,\,t=0$.The lines are the trajectories of $S_j$, and the red points are the endpoints of the stem which are expressed as Eqs.\ \eqref{eqc} and \eqref{eqd}.}\label{fig2-1}
\end{figure}

To ensure that $a_{12}\approx \infty$, it is necessary to set either $k_1\approx -k_2$ or $p_1\approx -p_2$. When $k_1\approx -k_2$, the table \ref{tab:t2} reveals that the amplitude of the stem $\widetilde{S_{1+2}}$ approaches zero, as illustrated in Fig.\ \ref{fig2-1} (a) and (b). Alternatively, if $p_1\approx -p_2$, the amplitude of $\widetilde{u_{1+2}}$ is almost zero, while $\widetilde{v_{1+2}}$ does not, as depicted in Fig.\ \ref{fig2-1} (c) and (d). Subsequently, an investigation of the 2-soliton \eqref{uv} is conducted on the cross-sectional curves situated on planes defined by $\xi_1+\xi_2+\ln a_{12}=0$, formulated as follows:
\begin{flalign}\label{cross23}
\begin{split}
\text{When $k_1\approx -k_2$,}&\\
u|_{\widetilde{l_{1+2}}}^{(1)}=&-\frac{2(k_1p_1+k_2p_2)(\e^{\theta_3}+\e^{\theta_4})+4g_1}{(\e^{\theta_3}+\e^{\theta_4}+2)^2};\\
v|_{\widetilde{l_{1+2}}}^{(1)}=&-\frac{2(k_1^2+k_2^2)(p_1-p_2)(\e^{\theta_3}+\e^{\theta_4})+4(k_1+k_2)(k_1p_1+k_2p_2)}{(p_1-p_2)(\e^{\theta_3}+\e^{\theta_4}+2)^2};
\end{split}
\end{flalign}
\begin{flalign}\label{cross24}
\begin{split}
\text{When $p_1\approx -p_2$,}&\\
u|_{\widetilde{l_{1+2}}}^{(2)}=&-\frac{2(k_1p_1+k_2p_2)(\e^{\theta_5}+\e^{\theta_6})+4g_1}{(\e^{\theta_5}+\e^{\theta_6}+2)^2};\\
v|_{\widetilde{l_{1+2}}}^{(2)}=&-\frac{2(k_1^2+k_2^2)(p_1-p_2)(\e^{\theta_5}+\e^{\theta_6})+4(k_1+k_2)(k_1p_1+k_2p_2)}{(p_1-p_2)(\e^{\theta_5}+\e^{\theta_6}+2)^2};
\end{split}
\end{flalign}
Here, $g_1,\,g_2$ are gien in \eqref{g12} and
$$\theta_3=\frac{p_1(k_2x-k_2^3t+\xi_2^0)-p_2(k_1x-k_1^3t+\xi_1^0+\ln a_{12})}{p_1+p_2},\,\theta_4=\frac{-p_1(k_2x-k_2^3t+\xi_2^0+\ln a_{12})+p_2(k_1x-k_1^3t+\xi_1^0)}{p_1+p_2},\,$$
$$\theta_5=\frac{k_1(p_2y-k_2^3t+\xi_2^0)-k_2(p_1y-k_1^3t+\xi_1^0+\ln a_{12})}{k_1+k_2},\,\theta_6=\frac{-k_1(p_2y-k_2^3t+\xi_2^0+\ln a_{12})+k_2(p_1y-k_1^3t+\xi_1^0)}{k_1+k_2}.$$
\begin{figure}[h!tb]
	\centering
	\raisebox{18ex}{(a)}\includegraphics[height=4cm,width=3.8cm]{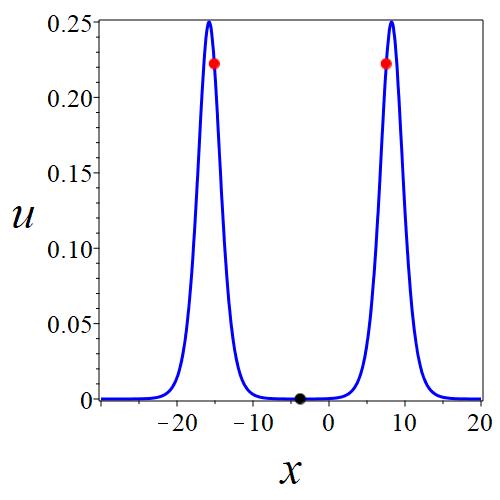}
	\raisebox{18ex}{(b)}\includegraphics[height=4cm,width=3.8cm]{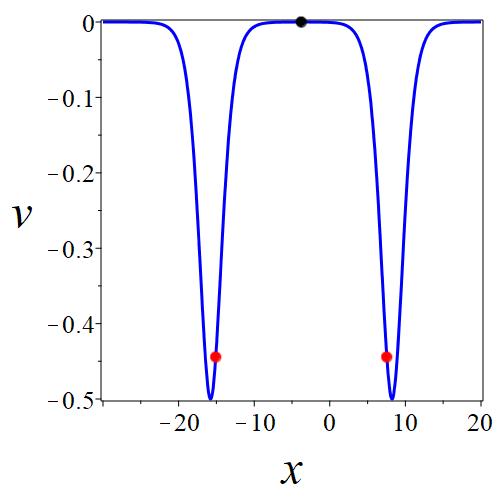}
    \raisebox{18ex}{(c)}\includegraphics[height=4cm,width=3.8cm]{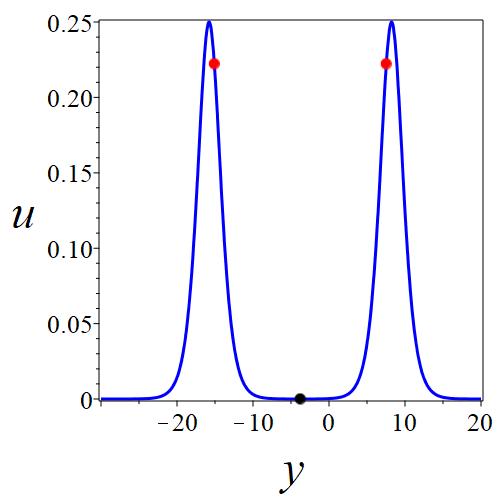}
	\raisebox{18ex}{(d)}\includegraphics[height=4cm,width=3.8cm]{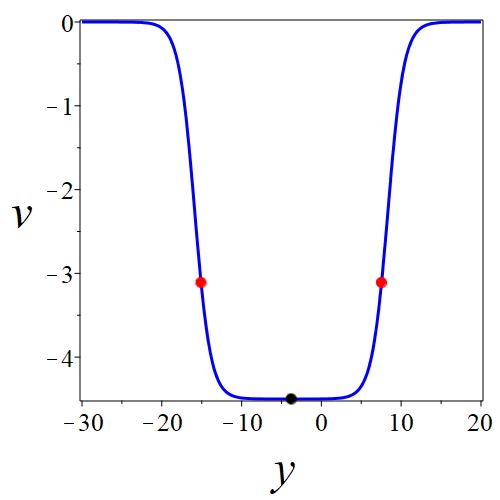}
	\caption{(a) (b) The cross-sectional curves \eqref{cross23} with parameters: $k_1=1,\,k_2=-1-\epsilon,\,p_1=1,\,p_2=2$; The black points are corresponding to $P_2$, and the red points are corresponding to $C$ and $D$. (c) (d) The cross-sectional curves \eqref{cross24} with parameters: $k_1=1,\,k_2=2,\,p_1=1,\,p_2=-1-10^{-10}$.
	}\label{fig2-2}
\end{figure}
\begin{figure}[h!tb]
	\centering
    \raisebox{18ex}{(a)}\includegraphics[height=4cm,width=3.8cm]{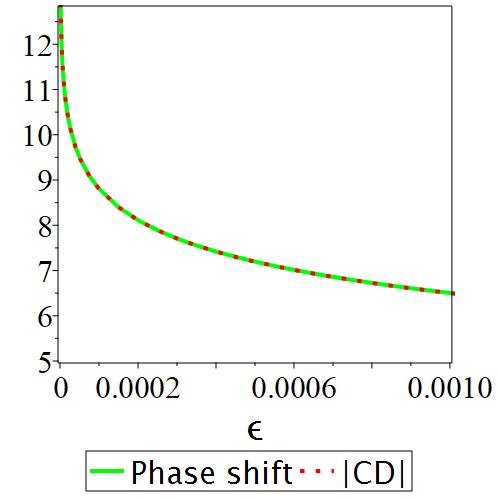}
	\raisebox{18ex}{(b)}\includegraphics[height=4cm,width=3.8cm]{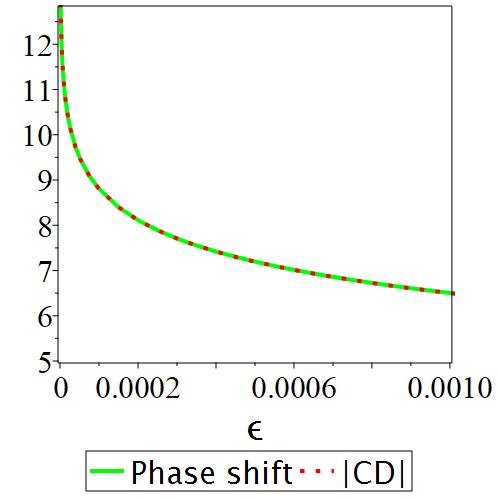}
	\caption{Parameters: (a) $k_1=1,\,k_2=-1-\epsilon,\,p_1=1,\,p_2=2$; (b) $k_1=1,\,k_2=2,\,p_1=1,\,p_2=-1-\epsilon$.
	}\label{fig2-3}
\end{figure}

Fig. \ref{fig2-2} illustrates the cross-sectional curves corresponding to Fig. \ref{fig2-1}. By taking the derivative of Eq.\ \eqref{cross23}, we can determine their extreme values. Notably, as the same to $S_{1-2}$, the constant length stem $S_{1+2}$ exhibits only one extreme point between $C$ and $D$. The coordinates of the extreme points of $u|_{\widetilde{l_{1+2}}}$ and $v|_{\widetilde{l_{1+2}}}$ in the $(x,,y)$-direction are as follows:
\begin{flalign}\label{extrem2}
\begin{split}
P_2:\,\left(\frac{(p_1-p_2)\ln a_{12}+2(k_1^3p_2-k_2^3p_1)t+2p_1\xi_2^0-2p_2\xi_1^0}{2(k_1p_2-k_2p_1)},\,-\frac{(k_1-k_2)\ln a_{12}+2(k_1^3k_2-k_1k_2^3)t+2k_1\xi_2^0-2k_2\xi_1^0}{2(k_1p_2-k_2p_1)}\right)
\end{split}
\end{flalign}

Remarkably, $P_2$ precisely corresponds to the midpoint of $CD$, depicted as black dots in Fig. \ref{fig2-2}. Substituting \eqref{extrem2} into \eqref{uv} and \eqref{2soliton} yields the extreme values as follows: $u(P_2)=-\frac{(k_1+k_2)(p_1+p_2)}{2}\sqrt{a_{12}}\approx -\frac{(k_1+k_2)(p_1+p_2)}{2}$ and $v(P_2)=-\frac{(k_1+k_2)^2}{2}\cdot\frac{(k_1+k_2)^2\sqrt{a_{12}}+(k_1-k_2)^2}{(k_1+k_2)^2(\sqrt{a_{12}}+1)}\approx -\frac{(k_1+k_2)^2}{2}$. These approximations further affirm the accuracy of the asymptotic forms \eqref{xasy06}. We can see again from Fig. \ref{fig2-2} that the two ends of the stem are not the maximum or minimum points of $u$ or $v$.

If we set $|k_1+k_2|=\epsilon\approx0$ or $|p_1+p_2|=\epsilon\approx0$ and $|k_j|,|p_j|\gg\epsilon$, we observe $|CD|\approx \Delta_{12}$. Fig. \ref{fig2-3} visually represents how both ``$|CD|$" and ``$|\Delta_{12}|$" change with $\epsilon$. Notably, when $\epsilon\approx0$, the $|CD|$ curve closely aligns with the phase curve. Furthermore, the smaller $\epsilon$ is, the bigger $|CD|$ is.
\section{Conclusions and discussions}\label{summary}
Although soliton theory has been developed in several decades since 1960s, the essential properties  of  the stem structures is still not an unravelled problem. In this study, we have undertaken a systematic examination of localized stem structures in two solitons for the ANNV system. The constant length stem structures are consequences of quasi-resonant collisions between two solitons, exhibiting distinct characteristics of spatial locality and temporal invariance. Specifically, our investigation delineates two discernible scenarios: one characterized by $a_{12}\approx 0$ indicating a weakly quasi-resonant collision, and the other by $a_{12}\approx +\infty$ signifying a strongly quasi-resonant collision.
\begin{itemize}
  \item Based on the asymptotic forms delineated in equations \eqref{xasy01}--\eqref{xasy03}, comprehensive insights into the trajectories, amplitude, and velocity of each soliton arm have been extracted and cataloged in Table \ref{tab:t1}, along with the corresponding relationships encapsulated in equations \eqref{uv01} and \eqref{l12}. It is notable that the termini of the stem structure are rigorously defined as the points of intersection between $l_1$ and $l_2$, as well as $\widetilde{l_1}$ and $\widetilde{l_2}$. Concomitantly, the length of the stem $S_{1-2}$ is explicitly formulated in Eq. \eqref{eqab}. Moreover, the parametric description of the stem's profile, articulated in equations \eqref{cross21} and \eqref{cross22}, has been scrutinized to unveil its critical extreme points, illustrated in Fig. \ref{fig1-2}. This indicates that the extremal curve of the stem is not a horizontal line like a genuine soliton but rather a curve centered at the midpoint of the stem, gradually attenuating on both sides. By setting $|k_1-k_2|=\epsilon\approx0$ or $|p_1-p_2|=\epsilon\approx0$ with $|k_j|,|p_j|\gg\epsilon$, it becomes apparent that as $\epsilon$ approaches zero, the proximity between the $|AB|$ curve and the phase curve intensifies, connoting that diminishing $\epsilon$ values correspond to elongated stems.
  \item Similarly, drawing upon the asymptotic forms expounded in equations \eqref{xasy04}--\eqref{xasy06}, an exhaustive analysis of the soliton arms' trajectories, amplitudes, and velocities has been compiled in Table \ref{tab:t1} and \ref{tab:t2}. Analogously, the terminal points of the stem structure are characterized as the intersections of $l_1$ with $\widetilde{l_2}$ and $\widetilde{l_1}$ with $l_2$. Analogous to the previous scenario, the length of the stem, denoted as $S_{1+2}$, is methodically defined in Eq. \eqref{eqcd}. Furthermore, the mathematical representation of the stem's profile, delineated in equations \eqref{cross23} and \eqref{cross24}, has undergone meticulous scrutiny to identify its pivotal extreme points, depicted in Fig. \ref{fig2-2}. It further suggests that the extremal curve of the stem is a smooth curve centered at its midpoint. Analogous to the prior analysis, the reduction of $|k_1+k_2|$ to $\epsilon\approx0$ or $|p_1+p_2|$ to $\epsilon\approx0$, accompanied by $|k_j|,|p_j|\gg\epsilon$, demonstrates the convergent alignment between the distance from $|CD|$ curve and the phase curve as $\epsilon$ tends to zero, signifying that diminishing $\epsilon$ values correspond to a lengthened stem structure.
\end{itemize}

A natural extension of our current work lies in the exploration of stem structures within higher order solitons. Beyond the quasi-resonant two-soliton case, the two-resonant three-soliton system can also give rise to localized structures, with the four surrounding branches interconnected by an intermediate soliton (stem structure) \cite{3soliton2r1,3soliton2r2}. In this phenomenon, the length of the stem gradually diminishes over time, leading to the fusion and subsequent separation of the surrounding soliton arms, while a new stem, oriented differently, reconnects them. This process, known as soliton reconnection, presents challenges in analyzing its local properties due to its complicated dynamic evolution over time. Future endeavors will focus on a comprehensive exploration of this phenomenon, aiming to provide deeper insights into the intricate dynamics of the ANNV system.


\vspace{\baselineskip} 
{{\bf Conflict statement}
	{The authors declare that they have no conflict of interests.}}

\vspace{\baselineskip} 
{{\bf Acknowledgments}
	{This work is supported by the National Natural Science Foundation of China (Grants 12071304 and 12201195), NUPTSF (Grants NY220161 and NY222169), the Natural Science Foundation of the Higher Education Institutions of Jiangsu Province (Grant 22KJB110004), Shenzhen Natural Science Fund (the Stable Support Plan Program) (Grant 20220809163103001), and Guangdong Basic and Applied Basic Research Foundation (Grant 2024A1515013106).}}
\section*{Refrence}
\addcontentsline{toc}{section}{Reference}

\end{document}